\documentstyle[aps,epsf,preprint]{revtex}
\begin{document}
\def\inseps#1#2{\def\epsfsize##1##2{#2##1} \centerline{\epsfbox{#1}}}
\def\top#1{\vskip #1\begin{picture}(290,80)(80,500)\thinlines \put(
65,500){\line(1, 0){255}}\put(320,500){\line(0, 1){
5}}\end{picture}}
\bibliographystyle{prsty}

\newcommand{\spmine}{1.1}
\newcommand{\myspace}{\edef\baselinestretch{\spmine}\Large\normalsize}
\myspace

\title{BERRY-PHASE AND SYMMETRY OF THE GROUND STATE IN DYNAMICAL JAHN-TELLER
SYSTEMS}
\author{Paolo De Los Rios}
\address{Max-Planck-Institut f\"ur Physik komplexer Systeme,
N\"othnitzer Str.\ 38, D-01187 Dresden, Germany}
\author{and Nicola Manini}
\address{European Synchrotron Radiation Facility, B.P. 220, F-38043
Grenoble C\'edex, France}

\date{\today}
\maketitle

\begin{abstract}
We show through selected examples, relevant to the physics of fullerene
ions, that the presence of a Berry phase in dynamical Jahn-Teller systems
does not guarantee the degeneracy of the ground state, contrary to what
previously believed.  Moreover, we introduce a new class of Jahn-Teller
systems with no Berry phase, as a generalization of the basic icosahedral
$H \otimes h$ case.
\end{abstract}


\section{Introduction}

The interest on degenerate electron-lattice interaction (Jahn-Teller
effect) in molecules and impurity centers in solids started in the $'30$s
and found large fields of application in the $'60$s. In recent years the
interest is coming back, excited by the realization of new systems of this
type calling for a revision of a number of commonly accepted beliefs.  One
of these systems, fullerene, is an highly symmetric icosahedral molecule:
this symmetry group implies large representations, thus large degeneracies
of the interacting electronic and vibrational states of the isolated ion.
New Jahn-Teller (JT) systems have therefore been studied, and they have
been shown to imply new exciting properties.  A particularly new and
surprising one has been demonstrated recently: the possibility of a change
in symmetry of the ground state.

The molecular symmetry reduction associated to the splitting of the
electronic state degeneracy is restored, as it is known for longtime, in
the vibronic ground state thanks to the dynamical JT effect (DJT), i.e.\
the coherent tunneling among equivalent distortions.  In all JT systems
known till few years ago, for single-electron occupancy, it was generally
accepted an empiric ``symmetry conservation rule'', sometimes known as
``Ham's theorem'', stating that the symmetry of the vibronic DJT ground
state, at any coupling strength, remains the same as that of the electronic
multiplet prior to coupling\cite{Bersuker}.  This phenomenon, not required
by any general constraint in the JT physics can be seen as the fingerprint
of a Berry phase in the entangled electronic-phononic dynamics, in turn an
apparently general feature.  New excitement sprung from the discovery of
the first dynamical JT system where the ground state is {\em not}
degenerate in the strong coupling limit, thus a system which is Berry-phase
free\cite{Delos96,Moate96}.  This is the model that in spherical symmetry
is indicated as $(L=2) \otimes (l=2)$, where electrons of angular momentum
$L=2$ interact with vibrations also belonging to an $l=2$ representation.
This system is relevant to the physics of fullerene ions C$_{60}^+$, where
the 5-fold degenerate electronic state has $H$ icosahedral label and the
quadrupolar distortions correspond to some of the $h$ modes\cite{Delos96}.
It has been shown both analytically and numerically that a nondegenerate
state in the vibronic spectrum moves down for increasing coupling, to cross
the 5-fold ground state at some finite value of the coupling parameter,
thus becoming the ground state at strong coupling.

In this paper we consider more in general the possibility of nondegenerate
vibronic ground states, and their compatibility with the presence of a
Berry phase in the coupled dynamics.

\section{Symmetry and topology considerations}

According to the general theory of the JT effect, an $N$-fold degenerate
electronic level corresponding to a representation $\Gamma$ of the symmetry
group of the molecule can interact with the vibrational modes corresponding
to representations $\{\Lambda\}$ contained in the symmetric part of $\Gamma
\otimes \Gamma$ (excluding the identical representation which is trivial).
The Hamiltonian for the case where exactly one mode $\omega_{\Lambda}$ of
each symmetry label $\Lambda$ is present and interacts linearly with
strength $g_{\Lambda}$ with the $N$-fold degenerate electronic level is
\begin{eqnarray}
H & = & \frac{1}{2} \sum_{\Lambda} \hbar \omega_{\Lambda}
\sum_{i=1}^{|\Lambda|} (p_{\Lambda i}^2 + q_{\Lambda i}^2) + \nonumber \\
& + &\frac 12
\sum_{\Lambda} g_{\Lambda} \sum_{i=1}^{|\Lambda|} \sum_{j,k=1}^{|\Gamma|}
q_{\Lambda i} c^\dagger_{\Gamma j} c_{\Gamma k}
<\Lambda i | \Gamma j \Gamma k>\;\;,
\label{hamiltonian}
\end{eqnarray}
where $|\Lambda|$ is the dimension of the $\Lambda$ representation (and
analogously for $|\Gamma|$) and $<\Lambda i | \Gamma j \Gamma k>$ are the
Clebsch-Gordan coefficients for the symmetry group $G$ of the molecule.  In
(\ref{hamiltonian}) we choose the real representation for the vibrational
degrees of freedom, and a second-quantized notation for the electrons.

In the special case where all frequencies $\omega_\Lambda $ and couplings
$g_\Lambda$ are equal, the symmetry group of (\ref{hamiltonian}) is raised
to $SO(N)$ and the problem reduces to a single-mode JT coupling between two
representations of that group\cite{Pooler80}.  Within this context it has
been shown\cite{Ceulemans87} that the set of minima of the Born-Oppenheimer
(BO) potential, corresponding to the most convenient classical distortions,
constitute a continuous manifold, referred to as Jahn-Teller manifold
(JTM).  In particular, due to the smooth adiabatic mapping from the
vibrational to the electronic sphere (meaning that to every point on the
JTM there corresponds a precise electronic eigenstate of the
electron-vibron interaction operator), the JTM shares the same symmetry and
topological properties of the electronic Hilbert space.  Since the
Hamiltonian for the JT problems we consider is real, the electronic space
is a sphere $S^{N-1}$ in the $N$-dimensional real vector space.  This
sphere has $SO(N)$ symmetry and, because of the mapping, the JTM is also
invariant for transformation under the group $SO(N)$, and not just $G$.
Moreover, the points on the electronic sphere are defined modulo a sign
(identification of the antipodes, representing the same electronic state).
Consequently, from the topological point of view, the electronic sphere is
a multiply connected manifold.  Due once more to the smooth mapping, the
JTM is also multiply connected.  When the system goes along a closed path
on the JTM and the electronic state follows adiabatically, it may happen
that the final electronic state differs from the initial state by a change
of sign.  In the most general case of a multiply connected JTM, there are
no smooth deformations of the closed path in vibrational space such that
this {\em anomaly} can be eliminated.  This is the essence of the Berry
phase in DJT systems: the electronic state, in the adiabatic approximation,
induces non trivial phases onto the vibrational dynamics.

\section{Results}

In the perspective of moving to the general case of different couplings
$g_\Lambda$'s, it is instructive to first consider the extreme case where
only one of the couplings $g_\Lambda$ is nonzero.  Problems of this kind
are, for example, $T \otimes t$ in cubic symmetry and $H \otimes h$ in
icosahedral symmetry.  In particular the latter case is of interest because
the icosahedral group is non simply reducible, so that there are two
different coupling schemes for the $H \otimes h$
problem\cite{Delos96,Moate96,Cesare87}.  While one of them is unique to the
discrete-group features of the icosahedral group, the other coupling may be
chosen to coincide with the $(L=2) \otimes (l=2)$ JT system introduced
above.

In this very special case, due to the symmetry properties of the
Clebsch-Gordan coefficients of $SO(3)$, even if the Hamiltonian has $SO(3)$
symmetry only, the JTM shares the same symmetry properties of the
electronic sphere, which here is the four-dimensional unit sphere $S^4$ in
the five-dimensional electronic space (with opposite points identified).
The JTM of the $(L=2) \otimes (l=2)$ system is therefore also a
four-dimensional manifold, immersed in the five-dimensional space of
distortions, enjoying complete $SO(5)$ symmetry.  The only possible
manifold satisfying these requirements, is isomorphic to another sphere
$S^4$.  Yet a sphere $S^4$ in a five dimensional space is a
simply-connected object, whereas the electronic sphere is not.  Therefore
some {\it pathology} is to be expected in the adiabatic mapping from the
vibrational space to the electronic one.  In particular, there must exist
at least one loop on the electronic sphere which maps to a single point in
vibrational space: under this condition, it is possible to drag and deform
smoothly any path connecting a point to its antipode on the electronic
sphere (corresponding to a closed path in vibrational space) to this {\it
pathological} path, which is equivalent to one point in the distortion
sphere.  This mechanism eliminates the Berry phase anomaly for this
problem.  As a consequence, as anticipated above, the strong-coupling
ground state is non degenerate.

The main reasons for the elimination of the Berry phase in the $(L=2)
\otimes (l=2)$ model are therefore the symmetry of the Clebsch-Gordan
coefficients and the exact matching of the dimensions of the electronic and
vibrational spaces: together they force the JTM to become a plain
(simply-connected) sphere.  In larger spaces of distortions, the JTM can be
both $SO(N)$-symmetric and multiply connected, and in all known cases it
really is.  This mechanism being clear, we predict that all the $SO(3)$
coupling schemes of the type $L \otimes l$, with even $l=L$ are DJT systems
with no Berry-phase entanglement.  In particular, the strong-coupling
ground state for this whole class of systems is non degenerate.  We
verified by numerical diagonalization that this is indeed the case for the
$(L=4) \otimes (l=4)$ case: an $L=0$ state originating (at weak coupling)
from the one-quantum multiplet, at $g_4\sim 8$ crosses down below the $L=4$
vibronic state which was the weak-coupling ground state.
Such kind of highly-symmetric systems may have applications for small to
medium-sized atomic clusters, or in odd-$A$ nuclei.

This result has a great relevance to the cases of coupling to a single
mode, but it is necessary to be careful in drawing general conclusions from
it.  One of the main results of this work is indeed that in a many-mode
situation, {\em the presence of a Berry phase does not automatically
guarantee the degeneracy of the ground state} of the system.

\begin{figure}
\epsfxsize 10.0cm
\inseps{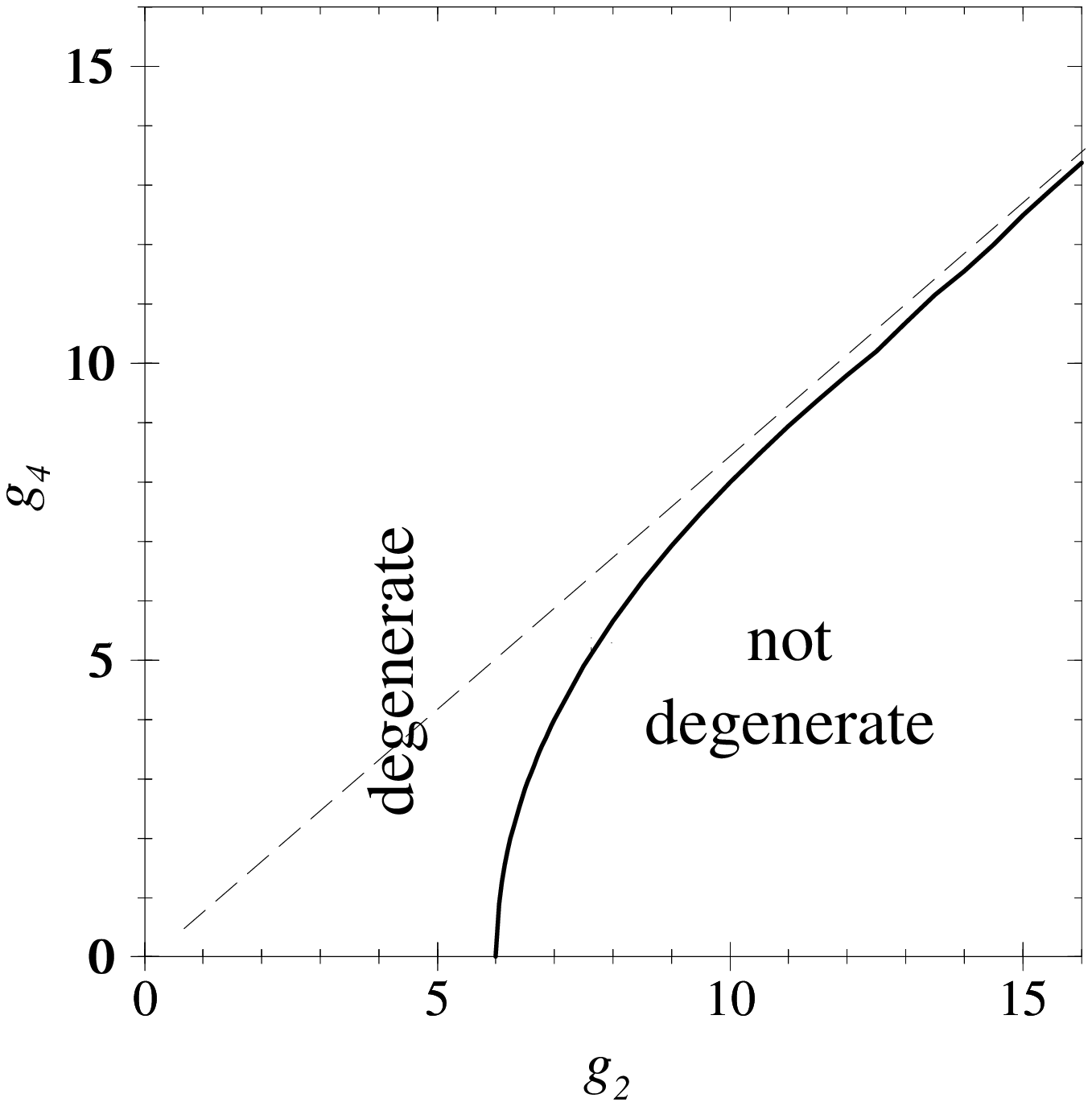}{0.8}
\caption{ The zero-temperature ``phase diagram'' of the $(L=2) \otimes (l=2
\oplus l=4)$ JT system in the space of the coupling parameters $g_2$ and
$g_4$, for fixed frequencies $\omega_2$ and $\omega_4$.  The solid line
marks the crossing between the five-fold (left) and the nondegenerate
(right) ground states.  The dashed line indicates the asymptotic behavior
of the crossover line at strong coupling.
\label{phased:fig}}
\end{figure}

This point is made very clear in the $(L=2) \otimes (l=2 \oplus l=4)$ case
(an $(L=2)$ electronic state linearly coupled to two spherical modes of
symmetry $l=2$ and $l=4$ respectively).  In that case, for equal
frequencies and couplings of the two modes, it was
shown\cite{Delos96,Pooler80} that the overall underlying symmetry of the
model is $SO(5)$.  For this special case, the presence of a Berry phase has
been explicitly demonstrated, as well as its consequences for the allowed
levels.  In the general case $g_2\neq g_4$, the symmetry is reduced to the
original $SO(3)$ and the large $SO(5)$ representations split into the
$SO(3)$ ones.  In particular, in the limit $g_4\to 0$, the vibronic
spectrum of $(L=2) \otimes (l=2)$ model is recovered.  For $g_4\equiv 0$
there is no Berry phase, and for $g_2\gtrsim 8$ the ground state is a
vibronic nondegenerate $L=0$ state separated by a finite energy gap from
the first $5$-fold degenerate $L=2$ excited state\cite{Delos96,Moate96}.
The continuous lowering of $g_4$ from $g_4=g_2$ to $g_4=0$ describes a
smooth mapping of a situation with a Berry phase entanglement (and a
degenerate ground state) to a case where this entanglement disappears
(nondegenerate ground state).  This implies a new case of level crossing,
at some intermediate value of $g_4$. In particular we have a crossover
value of $g_4$ for which the $L=0$ and $L=2$ states become degenerate.  At
strong coupling the energy gap $E[L=2]-E[L=0] = c_2 \omega_2 /g_2^2 +
O(g_2^{-4})$ for $g_4=0$ and $E[L=2]-E[L=0] = -c_4 \omega_4 /g_4^2 +
O(g_4^{-4})$ for $g_2=0$, where $c_2$ and $c_4$ are positive constants.
The curve of crossover points in the $(g_2,g_4)$ plane must get therefore
asymptotically close to the straight line $g_4/g_2=\sqrt{c_4\omega_4 /c_2
\omega_2}$ at strong coupling.  These considerations permit to draw the
zero-temperature ``phase diagram'' represented in Fig.~\ref{phased:fig}.

This intuitive picture captures the correct physics of the system, and it
is noteworth for describing a whole region of the phase diagram where the
presence of a Berry phase coexists with a nondegenerate $L=0$ ground state.
We need to reconcile the gradual, smooth effect of turning on the coupling
to the $ (l=4)$ mode, with the abrupt appearance of a Berry phase, which is
a topological effect, intrinsically non-perturbative, as soon as $g_4 \ne
0$.  In actuality, for any $g_4 \ne g_2$, the 30-fold degenerate
first-excited state (labeled [3,0] according to $SO(5)$) of the $g_4 = g_2$
``hypersymmetrycal'' spectrum is split\cite{Delos96} into its $L=0,3,4,6$
components ($SO(3)$ representations).  In particular the $L=0$ fragment is
the lowest one when $g_4/g_2<1$.  For small enough $g_4 /g_2$, this
nondegenerate state has the opportunity to localize as much as possible in
the (hyper)spherical potential well in the $l=2$ vibron space, therefore
becoming the ground state.  Even in this limit, however, the Berry phase
prescription is globally respected, since the $L=0$ (nondegenerate) ground
state is really a fragment of an odd ([3,0]) level selected by the Berry
phase: the ground state still fulfills the parity constraint imposed to the
low-energy $SO(5)$ representations by the Berry phase in the global $(l=2
\oplus l=4)$ space.  Note however that the same $L=0$ state, if seen
restricted to the $l=2$ vibration space, is naturally classified as an
[0,0] state for the symmetry group ($SO(5)$ again) of the JTM.

\section{Discussion}

Throughout our discussion we implicitly assumed a linear JT coupling scheme
as described by (\ref{hamiltonian}).  The introduction of quadratic terms
has usually effects similar to those produced by different linear coupling
$g_E\neq g_T$ in cubic symmetry\cite{ob69}, i.e.\ of ``warping'' the JTM.
The symmetry of the Hamiltonian, as a consequence, is reduced to the
symmetry group $G$ of the molecule, and so is that of the JTM.  Yet, the
properties of connectedness are topological properties, being therefore
robust against perturbations such as the warping: even if the symmetry is
reduced, the consequence of the presence/absence of the Berry phase on the
ground-state symmetry are unchanged.  Ham\cite{Ham87} has shown in
particular, for the $e \otimes E$ coupling scheme, that the introduction of
quadratic terms in the Hamiltonian, although splitting specific
degeneracies of excited states, does not substantially change the picture
as far as the Berry phase and the degeneracy of the ground state are
concerned.

In summary, we have shown that the presence/absence of the Berry phase
is not a sufficient criterion to decide whether the ground state is
degenerate or not at strong coupling.  However, the absence of a Berry
phase in at least one of the subsystem seems to be a necessary condition
for having a non-degenerate ground state.  We also illustrate the relevance
of the actual values of the coupling strengths between electrons and
vibrations, that only can really decide about the symmetry of the ground
state.  In this perspective, the experimental or {\em ab-initio}
determination of the actual values of such couplings is of the utmost
importance for practical systems such as those based on positive fullerene
ions.  Finally, we propose evidence for a whole family, following the
$(L=2) \otimes (l=2)$, of Berry-phase--free dynamical JT systems whose
strong-coupling ground state is, as a consequence, nondegenerate.

\section{Acknowledgements}

We are thankful to Erio Tosatti, Lu Yu, and Mary O'Brien for useful
discussion.

\noindent

\end{document}